\documentclass[showpacs,groupedaddress,secnumarabic,titlepage,nobibnotes,twocolumn,showkeys,aps]{revtex4}
\usepackage{makeidx}
\usepackage{amsmath}
\usepackage{eurosym}
\usepackage{amssymb}
\usepackage{graphicx}
\usepackage{dcolumn}
\usepackage{bm}
\usepackage{rotating}

\begin{document}

\title{Two sub-band conductivity of Si quantum well}
\author{M. Prunnila}
\thanks{Corresponding author \\
email: mika.prunnila@vtt.fi}
\author{J. Ahopelto}
\affiliation{VTT Information Technology, P.O.Box 1208, FIN-02044 VTT, Espoo, Finland }

\begin{abstract}
We report on two sub-band transport in double gate SiO$_{2}$-Si-SiO$_{2}$
quantum well with 14 nm thick Si layer at 270 mK. At symmetric well
potential the experimental sub-band spacing changes monotonically from 2.3
to 0.3 meV when the total electron density is adjusted by gate voltages
between $\sim 0.7\times 10^{16}$ $-3.0\times 10^{16}$ m$^{-2}$. The
conductivity is mapped in large gate bias window and it shows strong
non-monotonic features. At symmetric well potential and high density these
features are addressed to sub-band wave function delocalization in the
quantization direction and to different disorder of the top and bottom
interfaces of the Si well. Close to bi-layer/second sub-band threshold the
non-monotonic behavior is interpreted to arise from scattering from
localized band tail electrons.
\end{abstract}

\maketitle














\section{Introduction}

Developement of silicon-on-insulator technology has enabled fabrication of
silicon heterostructure devices where thin single crystalline Si film is
sandwitched between amorphous SiO$_{2}$ layers. This kind of SiO$_{2}$-Si-SiO%
$_{2}$ quantum well provides a unique material system where electron density
can be tuned in a broad range due to high potential barriers formed by the
SiO$_{2}$ layers. Previous work on SiO$_{2}$-Si-SiO$_{2}$ quantum wells has
mainly focused on single and bi-layer magneto transport \cite%
{takashina:2004b,prunnila:2005,prunnila:2005b}. In this work we focus on the
issue how the two sub-band or bi-layer transport affects the low temperature
conductivity of double gate SiO$_{2}$-Si-SiO$_{2}$ quantum well with 14 nm
thick Si layer.

\section{Experimental}

\begin{figure}[b]
\par
\begin{center}
\includegraphics[width=80mm,height=!]{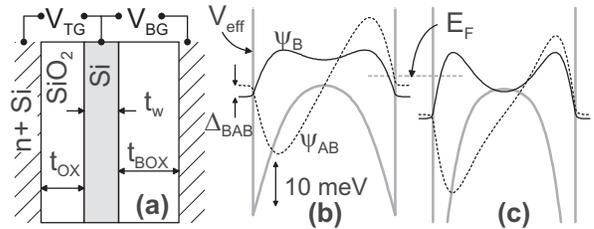}
\end{center}
\caption{(a) Schematic cross-section of the sample and gate biasing in the
experiments. The dimensions are $t_{\text{w}}$ =\ 14 nm, $t_{\text{OX}}$ =
40 nm and $t_{\text{BOX}}$ =\ 83 nm. (b)\&(c) Self-consistent bonding and
anti-bonding electron wave functions together with the effective potential $%
V_{\text{eff}}$ in the 14 nm thick Si well at balanced gate bias within the
Hartree approximation. The wave functions are offsetted so that $\Psi _{%
\text{B,AB}}$ = 0 is equal to the corresponding eigen energy $E_{\text{B,AB}%
}.$ (b): n = \ 1.0$\times $10$^{16}$ m$^{-2}.$ (c): n = \ 2.8$\times $10$%
^{16}$ m$^{-2}.$}
\label{wavefu}
\end{figure}

The samples were fabricated on commercially available (100)
silicon-on-insulator wafer as described in detail in Ref. \cite%
{prunnila:2005b}. The cross-sectional sample structure consist of n$^{+}$ Si
top gate, top gate oxide (OX), Si well, back gate oxide (BOX) and n$^{+}$ Si
back gate. The top gate is polycrystalline while the back gate is
crystalline silicon. All results reported here were obtained from a 100 $\mu 
$m wide Hall bar sample with 400 $\mu $m voltage probe distance. The layer
thickness for the Si well, top gate oxide and back gate oxide were $t_{\text{%
W}}$ = 14 nm, $t_{\text{OX}}$ = 40 nm, and $t_{\text{BOX}}$ = 83 nm,
respectively. Figure \ref{wavefu} shows the schematic device cross-section
together with self-consistent wave functions and effective potential at two
total electron density ($n$) values. 

In the experiments the sample was mounted to a sample holder of a He-3
cryostat, which was at base temperature (270 mK). The electron density was
determined from the Shubnikov- de Haas (SdH) oscillations of the diagonal
resistivity $\rho _{\text{xx}}$ utilizing the standard methods: in the
single sub-band gate bias windows $\rho _{\text{xx}}$ was measured as a
function of the gate voltages at constant magnetic field $B$ and $n$ was
determined from the minimum positions of $\rho _{\text{xx}}$. In the
presence of two sub-bands $\rho _{\text{xx}}$ was recorded as a function of $%
B$. Then a\ Fourier transform was performed numerically to $\rho _{\text{xx}%
}(1/B)$ and a peak position multiplied by $e/h$ ($e$ fundamental charge, $h$
Plank's constant) in the spectrum gave the sheet density per degeneracy of a
sub-band.

\section{Results and discussion}

\subsection{Sub-band densities at balanced gate bias}

\begin{figure}[t]
\par
\begin{center}
\includegraphics[width=80mm,height=!]{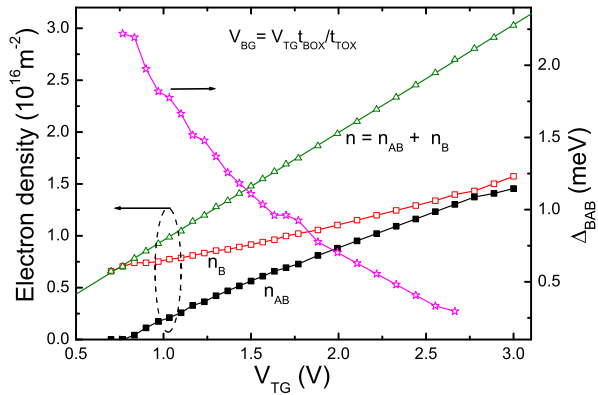}
\end{center}
\caption{Right axis: total electron density ($n$) and sub-band densities ($%
n_{\text{B}},n_{\text{AB}}$) as a function of top gate voltage along the
balanced gate line of a 14 nm thick Si well. The densities are determined
from SdH oscillations at 270 mK. Left axis: Sub-band energy spacing
calculated from $n_{\text{B}}$ and $n_{\text{AB}}$ assuming ideal 2D density
of states. }
\label{densities}
\end{figure}

Left vertical axis of Fig. \ref{densities} shows the different electron
densities as a function of top gate voltage $V_{\text{TG}}$ along the
balanced (or symmetric) gate bias line where the back gate voltage $V_{\text{%
BG}}=V_{\text{TG}}t_{\text{BOX}}/t_{\text{OX}}$. This choice of gate biases
produce symmetric potential in the Si well. The total density $n$ is
obtained by summing the bonding sub-band $n_{\text{B}}$ and anti-bonding
sub-band densities $n_{\text{AB}}$. On the balanced gate line the threshold
for $n_{\text{AB}}$ is at $V_{\text{TG}}\approx 0.8$ V and $n=n_{\text{B}%
}\approx 0.7\times 10^{16}$ m$^{-2}$. Note that the SdH\ method does not
reveal the degeneracies of the system. We have made an assumption that both
sub-bands arise from the high perpendicular mass valleys (non-primed
valleys), which gives the total degeneracy of \ four after including the
spin degeneracy. Our \textit{a priori} assumption is validated by noting
that $V_{\text{TG}}$ vs. $n=n_{\text{B}}+n_{\text{AB}}$ slope corresponds
accurately to the total gate capacitance. The light perpendicular mass
valleys (non-primed valleys) have a total degeneracy of eight and this
degeneracy would lead to incorrect total density. 

Right vertical axis of Fig. \ref{densities} shows the bonding anti-bonding
energy gap $\Delta _{\text{BAB}}$ which is obtained from $n_{\text{B}}$ and $%
n_{\text{AB}}$ by assuming ideal 2D density of states together with bulk
effective mass $m^{\ast }=m_{\text{t}}=0.19m_{0}$ parallel to the quantum
well plane. The increased gate drive and $n$ leads to expected reduction of $%
\Delta _{\text{BAB}}$ due to the potential barrier formation in the middle
of the Si well as demonstrated by the self-consistent calculations in Figs. %
\ref{wavefu}(b) and (c). The calculated $\Delta _{\text{BAB}}$ is larger
than the experimental one by $\sim 0.72$ meV (on average) in the density
range of Fig. \ref{densities} (not shown). This discrepancy can be explained
by noting that we have neglected the exchange and correlation effects in the
calculations.

\subsection{Conductivity and mobility}

\begin{figure}[t]
\par
\begin{center}
\includegraphics[width=0.98\linewidth]{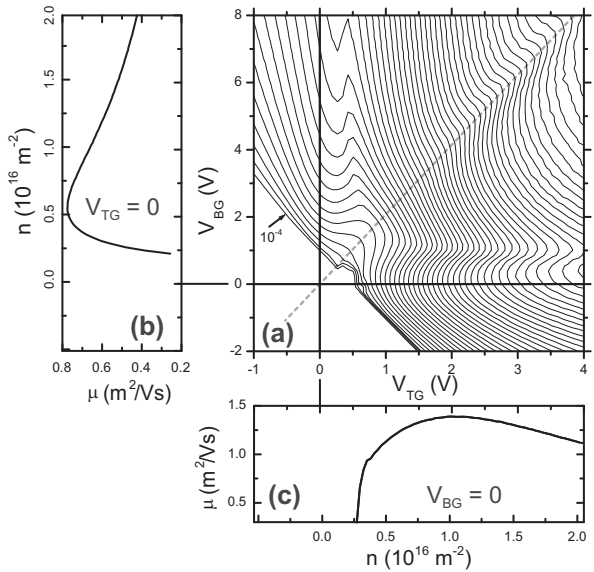}
\end{center}
\caption{(a) Contour plot of conductivity $\protect\sigma $ measured at 270
mK. The contour spacing is 10$^{-4}$ 1/$\Omega .$The dashed line is the
balanced gate line $V_{\text{BG}}=V_{\text{TG}}t_{\text{BOX}}/t_{\text{OX}}$%
. (b) Back interface mobility as a function of $V_{\text{BG}}$ (right
vertical axis) and $n$ (left vertical axis). (c) Top interface mobility as a
function of $V_{\text{TG}}$ and $n$.}
\label{Gmap}
\end{figure}

Figure \ref{Gmap}(a) shows\ (zero magnetic field) conductivity $\sigma
=1/\rho _{\text{xx}}$ measured as a function of top and back gate voltages 
\cite{note:arte2}. The scales of the voltage axes in Fig. \ref{Gmap}(a) are
chosen in such a manner that if we move perpendicularly to the balanced bias
line the electron density stays (roughly) constant, i.e., we move along $n=$%
const. line and merely shift the position of the electron gas inside the Si
slab. In the gate bias regions $V_{\text{TG}}\lesssim 0$ or $V_{\text{BG}%
}\lesssim 0$ $\ $only single sub-band is occupied and \ $\sigma $ behaves
monotonically, which is expected on the basis of simple Coulomb - surface
roughness scattering picture and is consistent with mobility measurements of
sub-10 nm thick Si well \cite{prunnila:2005b}.

The overall asymmetry of the conductivity with respect to the symmetric bias
line arises from the different quality of the top (Si-OX) and back (Si-BOX)
interfaces of the Si well. The back interface has substantially larger
disorder in comparison to the top interface as can be observed from \ Figs. %
\ref{Gmap}(b) and (c), which show the back interface and top interface
mobilities $\mu =\sigma /en$, i.e., the mobilities along the axes of Fig. %
\ref{Gmap}(a). Note that the $n$-axis of Figs. \ref{Gmap}(b) and (c)
,together with $n$ plotted in Fig. \ref{densities} ,\ also give an idea of
the magnitude of electron density. The observed disorder difference is
mainly due to larger surface roughness of the Si-BOX interface \cite%
{prunnila:2005b} consistent with the fact that the top-back mobility ratio
actually increases as a function of carrier density. At low density surface
roughness plays a minor role and the mobilities almost conincide, which is
also partly due to electron wave function spreading throughout the Si well.
The presence of the two Si-SiO$_{2}$ interfaces also reduce the peak
mobility and shift it towards higher electron densities \cite{prunnila:2004}%
. 

\begin{figure}[t]
\par
\begin{center}
\includegraphics[width=80mm,height=!]{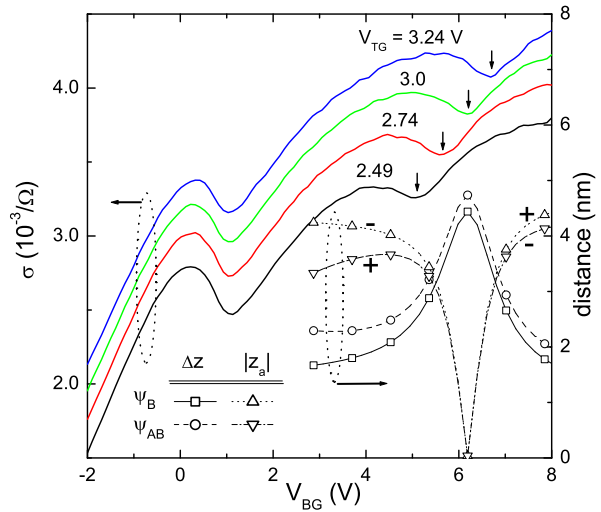}
\end{center}
\caption{Left axis: conductivity as a function of back gate voltage. $%
\downarrow $ indicate $V_{\text{BG}}=V_{\text{TG}}t_{\text{BOX}}/t_{\text{OX}%
}$. Right axis: deviation (position uncertainty) $\Delta z$ and absolute
value of average position $z_{a}$ of the sub-band wave functions. The zero
is in the middle of the well and +/- indicate the sign of $z_{a}$. The
calculation corresponds to the $V_{\text{TG}}$ =\ 3.0 V curve.}
\label{G(Vbg)}
\end{figure}

In the bias window $V_{\text{TG,BG}}\gtrsim 0$ the conductivity behaves
strongly non-monotonically. We can observe, e.g., that close to threshold $%
\sigma (V_{\text{TG}},V_{\text{BG}})|_{n=\text{const.}}$ has a local minimum
at symmetric bias. Note that at symmetric bias the second sub-band starts to
populate at $V_{\text{TG}}\approx 0.8$ V (as was shown in the previous
Sub-section) and the minimum is particularly strong when we are below this
threshold. If we increase the gate biases the minimum at $V_{\text{BG}}=V_{%
\text{TG}}t_{\text{BOX}}/t_{\text{OX}}$ for $\sigma (V_{\text{TG}},V_{\text{%
BG}})|_{n=\text{const.}}$ is clearly splitted into two local minima, which
follow closely the axes of the Figure. The minimum that follows $V_{\text{TG}%
}$-axis is illustrated more clearly in Fig. \ref{G(Vbg)} which shows $\sigma
(V_{\text{BG}},V_{\text{TG}}$=const.$).$ Note that if \ we would plot $%
\sigma (V_{\text{TG}},V_{\text{BG}}$=const.$)$ we would obtain a similar
curve (only the magnitude of $\sigma $ would be different due to different
quality of the Si-OX and Si-BOX interfaces). By comparing the high magnetic
field $\rho _{\text{xx}}$ data of Ref. \cite{prunnila:2005b} and Fig. \ref%
{Gmap}(a) we note that the local minimum in the vicinity of $\ V_{\text{BG,TG%
}}\sim 0$ for $\sigma (V_{\text{BG,TG}},V_{\text{TG,BG}}=$const.$)$ occurs
at the position which is the threshold where $\rho _{\text{xx}}$ starts to
show signatures of bi-layer transport.

The above non-monotonic behavior of \ $\sigma $ has a striking similarity to
the substrate bias experiments of bulk Si inversion layers, where it was
found that a positive substrate bias tends to reduce the mobility \cite%
{fowler:1975}. This effect has been addressed to spin flip scattering from
singly populated localized band tail states of higher sub-bands\cite%
{feng:1999} (, whose population increases with substrate bias in bulk
devices). Scattering from localized band tail electrons is the most probable
origin for the behavior of $\sigma $ in the vicinity of the $V_{\text{TG(BG)}%
}\sim 0$ at high $V_{\text{BG(TG)}}$ and around $V_{\text{BG}}=V_{\text{TG}%
}t_{\text{BOX}}/t_{\text{OX}}$ below the threshold of the second sub-band.
In detail, e.g., for the data in Fig. \ref{G(Vbg)}: the conductivity
saturates and starts to decrease because the population in the localized
band tail of the second sub-band increases and these localized electrons
scatter the electrons in the mobile first sub-band. Then at certain
threshold the Fermi level passes through "the mobility edge" of the second
sub-band and the conductivity starts to increase due to presence of two
sub-bands with extended states. 

If we increase the electron density by gate bias beyond the threshold of the
second sub-band the local minimum in $\sigma (V_{\text{BG}},V_{\text{TG}%
})|_{n=\text{const.}}$ at $V_{\text{BG}}=V_{\text{TG}}t_{\text{BOX}}/t_{%
\text{OX}}$ disappears due to presence of (solely) extended states. Then
above $V_{\text{TG}}\sim 2$ V another minimum appears. This is indicated by
the down arrows in Fig. \ref{G(Vbg)}. The origin of this feature is related
to the resonance effect observed in tunneling coupled double quantum wells
(often referred as resistance resonance) \cite%
{palevski:1990,ohno:1993,berk:1994}. At high electron density and gate bias
(at both gates) the double sub-band system in the Si well can be described
equivalently as a weak or medium coupling bi-layer, which is a direct
consequence of the barrier formation in the middle of the well [see Figs. %
\ref{wavefu}(b)\&(c)]. Far away from the balanced gate bias $V_{\text{BG}%
}=V_{\text{TG}}t_{\text{BOX}}/t_{\text{OX}}$ the barrier localizes the wave
functions of the different layers or sub-bands to the different sides of the
well. At balanced gate bias the sub-band wave functions delocalize across
the well. This localization - delocalization is illustrated in Fig. \ref%
{G(Vbg)} (right axis) where we show the average position $z_{a}=\left\langle
z\right\rangle =\left\langle \Psi \right\vert z\left\vert \Psi \right\rangle 
$ ($\Psi =\Psi _{B,AB}$) and deviation (position uncertainty) $\Delta z=%
\sqrt{\left\langle z^{2}\right\rangle -\left\langle z\right\rangle ^{2}}$ of
the sub-bands. At balanced bias $\Delta z$ has a maximum and $z_{a}=0$,
which results in mobility reduction of the sub-band that otherwise is
localized in the vicinity of the less disordered interface while the
mobility of the other sub-band stays practically unaffected \cite{ohno:1993}%
. This leads to the local minimum in $\sigma $ at $V_{\text{BG}}=V_{\text{TG}%
}t_{\text{BOX}}/t_{\text{OX}}$ at high electron densities$.$ The detailed
form of this minimum is strongly affected by the momentum relaxation time
and quantum lifetime \cite{berk:1994} and further discussion will be
published elsewhere.

\section{Summary}

We have reported on two sub-band transport in 14 nm thick Si quantum well at
270 mK. The conductivity of the quantum well showed non-monotonic double
gate bias dependency. At symmetric well potential and high density these
were addressed to sub-band wave function delocalization in the quantization
direction and to different disorder of the top and bottom interfaces of the
Si well. In the gate bias regimes close to 2$^{\text{nd}}$ sub-band /
bi-layer threshold the non-monotonic behavior was interpreted to arise from
spin flip scattering from singly populated localized band tail states of
higher sub-bands.

\begin{acknowledgments}
M. Markkanen is thanked for assistance in the sample fabrication. Academy of
Finland is acknowledged for financial support through project \# 205467.
\end{acknowledgments}

\bibliographystyle{apsrev}
\bibliography{twodeg_refs}

\begin{thebibliography}{10}
\expandafter\ifx\csname natexlab\endcsname\relax\def\natexlab#1{#1}\fi
\expandafter\ifx\csname bibnamefont\endcsname\relax
  \def\bibnamefont#1{#1}\fi
\expandafter\ifx\csname bibfnamefont\endcsname\relax
  \def\bibfnamefont#1{#1}\fi
\expandafter\ifx\csname citenamefont\endcsname\relax
  \def\citenamefont#1{#1}\fi
\expandafter\ifx\csname url\endcsname\relax
  \def\url#1{\texttt{#1}}\fi
\expandafter\ifx\csname urlprefix\endcsname\relax\def\urlprefix{URL }\fi
\providecommand{\bibinfo}[2]{#2}
\providecommand{\eprint}[2][]{\url{#2}}

\bibitem[{\citenamefont{Takashina et~al.}(2004)\citenamefont{Takashina,
  Fujiwara, Horiguchi, Takahashi, and Hirayama}}]{takashina:2004b}
\bibinfo{author}{\bibfnamefont{K.}~\bibnamefont{Takashina}},
  \bibinfo{author}{\bibfnamefont{A.}~\bibnamefont{Fujiwara}},
  \bibinfo{author}{\bibfnamefont{S.}~\bibnamefont{Horiguchi}},
  \bibinfo{author}{\bibfnamefont{Y.}~\bibnamefont{Takahashi}},
  \bibnamefont{and} \bibinfo{author}{\bibfnamefont{Y.}~\bibnamefont{Hirayama}},
  \bibinfo{journal}{Phys. Rev. B} \textbf{\bibinfo{volume}{69}},
  \bibinfo{pages}{161304(R)} (\bibinfo{year}{2004}).

\bibitem[{\citenamefont{Prunnila
  et~al.}(2005{\natexlab{a}})\citenamefont{Prunnila, Ahopelto, and
  Sakaki}}]{prunnila:2005}
\bibinfo{author}{\bibfnamefont{M.}~\bibnamefont{Prunnila}},
  \bibinfo{author}{\bibfnamefont{J.}~\bibnamefont{Ahopelto}}, \bibnamefont{and}
  \bibinfo{author}{\bibfnamefont{H.}~\bibnamefont{Sakaki}},
  \bibinfo{journal}{Phys. Stat. Sol.(a)} \textbf{\bibinfo{volume}{202}},
  \bibinfo{pages}{970} (\bibinfo{year}{2005}{\natexlab{a}}).

\bibitem[{\citenamefont{Prunnila
  et~al.}(2005{\natexlab{b}})\citenamefont{Prunnila, Ahopelto, and
  Gamiz}}]{prunnila:2005b}
\bibinfo{author}{\bibfnamefont{M.}~\bibnamefont{Prunnila}},
  \bibinfo{author}{\bibfnamefont{J.}~\bibnamefont{Ahopelto}}, \bibnamefont{and}
  \bibinfo{author}{\bibfnamefont{F.}~\bibnamefont{Gamiz}}, \bibinfo{howpublished}{Solid
  State Electronics, in print, also available at
  http://lanl.arxiv.org/cond-mat/0506073} (\bibinfo{year}{2005}{\natexlab{b}}).

\bibitem[{not()}]{note:arte2}
\bibinfo{howpublished}{The obscured behavior of the $\sigma$ contours in the
  vicinity of the threshold and below the symmetric bias line is an
  experimental artefact explained in \cite{prunnila:2005b}.}

\bibitem[{\citenamefont{Prunnila et~al.}(2004)\citenamefont{Prunnila, Ahopelto,
  and Gamiz}}]{prunnila:2004}
\bibinfo{author}{\bibfnamefont{M.}~\bibnamefont{Prunnila}},
  \bibinfo{author}{\bibfnamefont{J.}~\bibnamefont{Ahopelto}}, \bibnamefont{and}
  \bibinfo{author}{\bibfnamefont{F.}~\bibnamefont{Gamiz}},
  \bibinfo{journal}{Appl. Phys. Lett.} \textbf{\bibinfo{volume}{84}},
  \bibinfo{pages}{2298} (\bibinfo{year}{2004}).

\bibitem[{\citenamefont{Fowler}(1975)}]{fowler:1975}
\bibinfo{author}{\bibfnamefont{A.~B.} \bibnamefont{Fowler}},
  \bibinfo{journal}{Phys. Rev. Lett.} \textbf{\bibinfo{volume}{34}},
  \bibinfo{pages}{15} (\bibinfo{year}{1975}).

\bibitem[{\citenamefont{Feng et~al.}(1999)\citenamefont{Feng, Popovic, and
  Washburn}}]{feng:1999}
\bibinfo{author}{\bibfnamefont{X.~G.} \bibnamefont{Feng}},
  \bibinfo{author}{\bibfnamefont{D.}~\bibnamefont{Popovic}}, \bibnamefont{and}
  \bibinfo{author}{\bibfnamefont{S.}~\bibnamefont{Washburn}},
  \bibinfo{journal}{Phys. Rev. Lett.} \textbf{\bibinfo{volume}{83}},
  \bibinfo{pages}{368} (\bibinfo{year}{1999}).

\bibitem[{\citenamefont{Palevski et~al.}(1990)\citenamefont{Palevski, Beltram,
  Capasso, Pfeiffer, and West}}]{palevski:1990}
\bibinfo{author}{\bibfnamefont{A.}~\bibnamefont{Palevski}},
  \bibinfo{author}{\bibfnamefont{F.}~\bibnamefont{Beltram}},
  \bibinfo{author}{\bibfnamefont{F.}~\bibnamefont{Capasso}},
  \bibinfo{author}{\bibfnamefont{L.}~\bibnamefont{Pfeiffer}}, \bibnamefont{and}
  \bibinfo{author}{\bibfnamefont{K.~W.} \bibnamefont{West}},
  \bibinfo{journal}{Phys. Rev. Lett.} \textbf{\bibinfo{volume}{65}},
  \bibinfo{pages}{1929} (\bibinfo{year}{1990}).

\bibitem[{\citenamefont{Ohno et~al.}(1993)\citenamefont{Ohno, Tsuchiya, and
  Sakaki}}]{ohno:1993}
\bibinfo{author}{\bibfnamefont{Y.}~\bibnamefont{Ohno}},
  \bibinfo{author}{\bibfnamefont{M.}~\bibnamefont{Tsuchiya}}, \bibnamefont{and}
  \bibinfo{author}{\bibfnamefont{H.}~\bibnamefont{Sakaki}},
  \bibinfo{journal}{Appl. Phys. Lett.} \textbf{\bibinfo{volume}{62}},
  \bibinfo{pages}{1952} (\bibinfo{year}{1993}).

\bibitem[{\citenamefont{Berk et~al.}(1994)\citenamefont{Berk, Kamenev,
  Palevski, Pfeiffer, and West}}]{berk:1994}
\bibinfo{author}{\bibfnamefont{Y.}~\bibnamefont{Berk}},
  \bibinfo{author}{\bibfnamefont{A.}~\bibnamefont{Kamenev}},
  \bibinfo{author}{\bibfnamefont{A.}~\bibnamefont{Palevski}},
  \bibinfo{author}{\bibfnamefont{L.~N.} \bibnamefont{Pfeiffer}},
  \bibnamefont{and} \bibinfo{author}{\bibfnamefont{K.~W.} \bibnamefont{West}},
  \bibinfo{journal}{Phys. Rev. B.} \textbf{\bibinfo{volume}{50}},
  \bibinfo{pages}{15420} (\bibinfo{year}{1994}).

\end{thebibliography}

\end{document}